# Cavity quantum electrodynamics with ferromagnetic magnons in a small yttrium-iron-garnet sphere

Dengke Zhang,[1,†] Xin-Ming Wang,[2,†] Tie-Fu Li,[3,1,*] Xiao-Qing Luo,[1] Weidong Wu,[2] Franco Nori,[4,5] and J. Q. You[1,*]

[1]*Quantum Physics and Quantum Information Division, Beijing Computational Science Research Center, Beijing 100084, China*

[2]*Research Center of Laser Fusion, China Academy of Engineering Physics, P.O. Box 919-987, Mianyang 621900, China*

[3]*Institute of Microelectronics, Tsinghua University, Beijing 100084, China*

[4]*Center for Emergent Matter Science, RIKEN, Wako-shi 351-0198, Japan*

[5]*Physics Department, The University of Michigan, Ann Arbor, MI 48109-1040, USA*

†These authors contributed equally to this work.

*Correspondence and requests for materials should be addressed to T.F.L. (litf@tsinghua.edu.cn) or J.Q.Y. (jqyou@csrc.ac.cn).

Hybridizing collective spin excitations and a cavity with high cooperativity provides a new research subject in the field of cavity quantum electrodynamics and can also have potential applications to quantum information. Here we report an experimental study of cavity quantum electrodynamics with ferromagnetic magnons in a small yttrium-iron-garnet (YIG) sphere at both cryogenic and room temperatures. We observe for the first time a strong coupling of the same cavity mode to both a ferromagnetic-resonance (FMR) mode and a magnetostatic (MS) mode near FMR in the quantum limit. This is achieved at a temperature ~ 22 mK, where the average microwave photon number in the cavity is less than one. At room temperature, we also observe strong coupling of the cavity mode to the FMR mode in the same YIG sphere and find a slight increase of the damping rate of the FMR mode. These observations reveal the extraordinary robustness of the FMR mode against temperature. However, the MS mode becomes unobservable at room temperature in the measured transmission spectrum of the microwave cavity containing the YIG sphere. Our numerical simulations show that this is due to a drastic increase of the damping rate of the MS mode.

## Introduction

Hybrid quantum circuits combining two or more physical systems can harness the distinct advantages of different physical systems to better explore new phenomena and potentially bring about novel quantum technologies (for a review, see ref. 1). Among them, a hybrid system consisting of a coplanar waveguide resonator and a spin ensemble was proposed[2] and experimentally utilized[3-8] to implement both on-chip cavity quantum electrodynamics and quantum information processing. This spin ensemble is usually based on dilute paramagnetic impurities, such as nitrogen-vacancy (NV) centers in diamond[9,10] and rare-earth ions doped in a crystal[11]. By increasing the density of the paramagnetic impurities, strong and even ultrastrong couplings between the cavity and the spin ensemble can be achieved, but the coherence time of the spin excitations is drastically shortened. Indeed, it is a challenging task to realize both good quantum coherence of the spin ensemble and its strong coupling to a cavity.

Very recently, collective spins in a yttrium-iron-garnet (YIG) ferromagnetic material were explored to achieve their strong[12-14] and even ultrastrong couplings[15] to a microwave cavity. In contrast to spin ensembles based on dilute paramagnetic impurities, these spins[12] are strongly exchange-coupled and have a much higher density ($\sim 4.2 \times 10^{21}$ cm$^{-3}$). Because of this high spin density, a strong coupling of the spin excitations to the cavity can be easily realized using a YIG sample as small as sub-millimeter in size. Also, the ultrastrong coupling regime becomes readily reachable by either increasing the size of the YIG sample or using a specially-designed microwave cavity[15]. Moreover, the contribution of magnetic dipole interactions to the linewidth of spin excitations, which can play a dominant role among paramagnetic impurities, is suppressed by the strong exchange coupling between the ferromagnetic electrons[16]. Thus, when the same spin density is involved, the spin excitations in YIG can exhibit much better quantum coherence than those of the paramagnetic impurities.

The YIG material is ferromagnetic at both cryogenic and room temperatures because its Curie temperature is as high as 559 K. Indeed, a strong coupling between a microwave cavity and the collective spin excitations related to the ferromagnetic resonance (FMR) was observed in different YIG samples at either cryogenic[12, 13, 15] or room temperature[14]. Here we report a direct observation of the strong coupling between FMR magnons and microwave photons at both cryogenic and room temperatures by using the same small YIG sphere in a three-dimensional (3D) cavity. This allows us to directly compare the quantum coherence of the same FMR mode at both cryogenic and room temperatures. We explain why the FMR spin excitations can be described as non-interacting magnons even at room temperature and its extraordinary robustness against temperature. Moreover, at cryogenic temperatures, we have also observed a strong coupling of the microwave photons to another collective spin mode, i.e., a magnetostatic (MS) mode, near the FMR. We find that this newly observed collective mode of spins exhibits quantum coherence nearly as good as the FMR mode. In addition, this collective spin mode can also be described as non-interacting magnons at room temperature, but becomes unobservable due to the drastic decrease of its quantum coherence. In short, compared to very recent studies[12-15], our experiment demonstrates for the first time the experimentally-achieved strong coupling of both FMR and MS modes to the same cavity mode and unveils quantum-coherence properties of the ferromagnetic magnons at both cryogenic and room temperatures.

**Results**

**Experimental setup and model Hamiltonian**
We illustrate the experimental setup in Fig. 1. A small YIG sphere with a diameter of 0.32 mm is mounted in a 3D rectangular microwave cavity with dimensions $50 \times 18 \times 3$ mm$^3$ (see Fig. 1a). The frequency of the fundamental cavity mode TE$_{101}$ is measured to be $\omega_{c,1}/2\pi = 8.855$ GHz and the frequency of the second cavity mode TE$_{102}$ is $\omega_{c,2}/2\pi = 10.306$ GHz (see Supplementary Material). By adjusting the lengths of the pins inside the input and output ports, the coupling rates related to these ports are tuned to $\kappa_{i,1}/2\pi = 0.19$ MHz and $\kappa_{o,1}/2\pi = 0.20$ MHz for mode TE$_{101}$, and $\kappa_{i,2}/2\pi = 0.85$ MHz and $\kappa_{o,2}/2\pi = 0.99$ MHz for mode TE$_{102}$. To achieve strong couplings between the YIG sample and these two cavity modes, we place the YIG sphere at the center of one short edge of the rectangular cavity where both cavity modes, TE$_{101}$ and TE$_{102}$, have stronger magnetic fields parallel to the short edge (see Figs. 1b and 1c). In our experiment, the YIG sphere is mounted, so that the microwave magnetic fields of the cavity modes TE$_{101}$ and TE$_{102}$ at the YIG sphere are nearly parallel to the crystalline axis $\langle 110 \rangle$. Also, a static magnetic field perpendicular to the microwave magnetic field and parallel to the crystalline axis $\langle 100 \rangle$ of the YIG sphere is applied. This static magnetic field can be tuned to drive the magnons in resonance with the cavity modes TE$_{101}$ and TE$_{102}$, respectively.

For a small YIG sample embedded in a microwave cavity, we observed a strong coupling between the fundamental cavity mode and the FMR mode, i.e., the Kittel mode (see Fig. 2a), as also reported in refs. 12-15. This FMR corresponds to a collective mode of spins with zero wavevector (i.e., in the long-wavelength limit) at which all exchange-coupled spins uniformly precess in phase together. In addition to this FMR, there are other long-wavelength collective modes of spins called the magnetostatic (MS) modes[17, 18]. These MS modes are also called dipolar spin waves under magnetostatic approximation, where the magnetic dipolar interactions dominate both the electric and exchange interactions[19]. In this condition, the wave number $k_{MS}$ of a MS mode satisfies $k_0 \ll k_{MS} \ll \sqrt{1/\Lambda_{ex}}$, where $k_0 = \omega\sqrt{\mu_0\varepsilon}$ is the wave number of the microwave field propagating in the YIG material and the exchange constant is $\Lambda_{ex} = 3 \times 10^{-16}$ m$^2$. Meanwhile, the microwave magnetic field $\mathbf{h}$ related to a MS mode of the YIG sphere satisfies the magnetostatic equation[19]

$$\nabla \times \mathbf{h} = \frac{\partial \mathbf{D}}{\partial t} = -\frac{k_0^2 \mathbf{k}_{MS} \times \mathbf{m}}{k_0^2 - k_{MS}^2} \approx 0, \quad (1)$$

where $\mathbf{D}$ is the microwave electric displacement vector and the magnetization $\mathbf{m}$ is excited by $\mathbf{h}$. The right side of equation (1) is approximately zero because $k_{MS} \gg k_0$, and the equation $\partial \mathbf{D}/\partial t \approx 0$ implies that the MS modes are essentially static. Also, like the FMR mode, the MS modes are discrete modes constrained by the boundary condition of the YIG sphere. Because the magnetic dipolar interactions dominate the exchange interactions, these modes are also called rigid discrete modes. However, unlike the FMR mode with uniform precession, the MS modes are non-uniform precession modes holding inhomogeneous magnetization and have a spatial variation comparable to the sample dimensions[17-19].

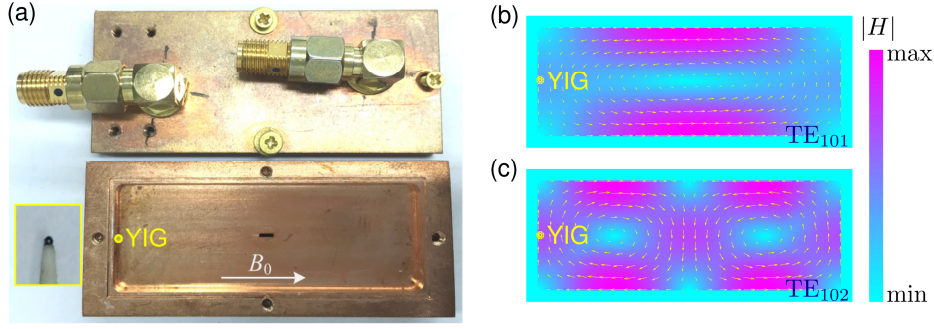

**Figure 1 | A high-finesse 3D microwave cavity containing a small YIG sphere.** (a) The experimentally used rectangular 3D cavity is made of oxygen-free copper and has dimensions $50 \times 18 \times 3$ mm$^3$. The upper panel shows the top cover of the cavity, where two connectors are attached for microwave transmission. In the lower panel, a small YIG sphere with a diameter of 0.32 mm is mounted at the center of one short edge of the rectangular cavity. The inset at the left is a magnified picture of the YIG sphere. (b) Simulated magnetic-field distribution of the fundamental cavity mode TE$_{101}$. (c) Simulated magnetic-field distribution of the second cavity mode TE$_{102}$. A static magnetic field $B_0$ is applied parallel to the long edge of the cavity, and the microwave magnetic field at the YIG sphere is perpendicular to the static magnetic field.

Here we utilize the transmission spectrum of the microwave cavity to demonstrate for the first time the experimentally-achieved strong coupling of both FMR and MS collective modes to the same mode (either the fundamental or second mode) of the 3D rectangular cavity. Because the FMR and MS modes are long-wavelength rigid discrete modes of spins in the YIG sphere, the exchange interactions between electron spins can be neglected[18, 20]. Thus, similar to the model Hamiltonian in refs. 21 and 22, when the FMR and a MS mode are both involved, the YIG-cavity Hamiltonian can be written as (setting $\hbar = 1$)

$$\begin{aligned} H &= \omega_c a^\dagger a + g\mu_B B_z^{(FMR)} S_z^{(FMR)} + g_{FMR}\left[aS_+^{(FMR)} + a^\dagger S_-^{(FMR)}\right] \\ &+ g\mu_B B_z^{(MS)} S_z^{(MS)} + g_{MS}\left[aS_+^{(MS)} + a^\dagger S_-^{(MS)}\right], \end{aligned} \quad (2)$$

where $\omega_c$ is the angular frequency of the cavity mode considered (either TE$_{101}$ or TE$_{102}$), $g$ the electron

$g$-factor, and $\mu_B$ the Bohr magneton; $B_z^{(\mathrm{FMR})}$ and $B_z^{(\mathrm{MS})}$ are the effective magnetic fields experienced, respectively, by the FMR and MS modes of the YIG sphere; while $a^\dagger$ ($a$) is the photon creation (annihilation) operator and $g_{\mathrm{FMR(MS)}}$ is the coupling strength of the cavity mode to a single spin in the FMR (MS) mode. Because the frequencies of different cavity modes are away from each other, we can then write a separate Hamiltonian for each cavity mode. The collective spin operators are given by $\mathbf{S}^{(m)} = (S_x^{(m)}, S_y^{(m)}, S_z^{(m)})$, where $S_\pm^{(m)} = S_x^{(m)} \pm i S_y^{(m)}$, and $m = \mathrm{FMR\,(MS)}$ denotes the FMR (MS) mode. These collective spin operators are related to the magnon operators via the Holstein-Primakoff transformation: $S_+^{(m)} = b_m^\dagger(\sqrt{2S^{(m)} - b_m^\dagger b_m})$, $S_-^{(m)} = (\sqrt{2S^{(m)} - b_m^\dagger b_m})b_m$, and $S_z^{(m)} = b_m^\dagger b_m - S^{(m)}$, where $S^{(m)}$ is the total spin number of the corresponding collective spin operator. For the low-lying excitations with $\langle b_m^\dagger b_m \rangle / 2S^{(m)} \ll 1$, one has $S_+^{(m)} \approx b_m^\dagger \sqrt{2S^{(m)}}$, and $S_-^{(m)} \approx b_m \sqrt{2S^{(m)}}$, i.e., the collective spin excitations can be described as non-interacting magnons. Then, the Hamiltonian (2) is reduced to

$$H = \omega_c a^\dagger a + \sum_{m=\mathrm{FMR,MS}} \left[\omega_m b_m^\dagger b_m + \tilde{g}_m (a b_m^\dagger + a^\dagger b_m)\right], \quad (3)$$

where $\omega_m = g\mu_B B_z^{(m)}$ is the angular frequency of the magnon mode (either FMR or MS mode) and $\tilde{g}_m = g_m \sqrt{2S^{(m)}}$ is the corresponding magnon-photon coupling strength. By diagonalizing the Hamiltonian in equation (3), we obtain the energy levels of the magnon-polariton (see the green curves in Fig. 2).

**Experimental measurements at cryogenic temperature**

An avoided crossing occurs around $\omega_c = \omega_{\mathrm{FMR}}$ ($\omega_{\mathrm{MS}}$) because of the strong coupling between the cavity mode and the FMR (MS) mode. Thus, three magnon-polariton branches appear (see the green curves in Fig. 2a). By fitting the experimental results in Fig. 2a with equation (3), we find that $\tilde{g}_{\mathrm{FMR}}/2\pi = 5.4$ MHz and $\tilde{g}_{\mathrm{MS}}/2\pi = 1.4$ MHz, for the case of the fundamental cavity mode $TE_{101}$ coupled to the FMR and MS modes, respectively. However, when the second cavity mode $TE_{102}$ interacts with the FMR and MS modes, the corresponding coupling strengths increase to $\tilde{g}_{FMR}/2\pi = 7.5$ MHz and $\tilde{g}_{MS}/2\pi = 8.3$ MHz (see Fig. 2b). It can be seen that the coupling strength between the cavity mode and the FMR mode does not change much when shifting the cavity mode from $TE_{101}$ to $TE_{102}$, but the coupling strength between the cavity mode and the MS mode changes drastically from $TE_{101}$ to $TE_{102}$. Surprisingly, $\tilde{g}_{MS}$ is even stronger than $\tilde{g}_{FMR}$ when the cavity mode is $TE_{102}$. This is because of a better overlap between the MS mode and the second cavity mode. In YIG, the only magnetic ions are the ferric ions with spin number $s = 5/2$. Since the FMR mode involves uniform, in-phase precession of all spins in the YIG sphere, one has $S^{(FMR)} = Ns$, where $N$ is the total number of spins. Therefore, it is natural to obtain $\tilde{g}_{FMR} = g_{FMR}\sqrt{5N}$. The coupling strength of a single spin to the cavity mode can be calculated by[14] $g_{FMR} = \eta \gamma_e \sqrt{\hbar \omega_c \mu_0 / V_c}/2$, where $V_c$ is the volume of the cavity mode with frequency $\omega_c$, $\gamma_e = 2\pi \times 28.0$ GHz/T is the gyromagnetic ratio, and the overlap coefficient $\eta$ describes the spatial overlap and polarization-matching conditions between the cavity and magnon modes. With our experimental parameters, the calculated $g_{FMR}/2\pi$ is 15.8 mHz for the cavity mode $TE_{101}$. Thus, from the experimentally obtained $\tilde{g}_{FMR}/2\pi = 5.4$ MHz, the total number of spins is estimated to be $N \sim 2.3 \times 10^{16}$.

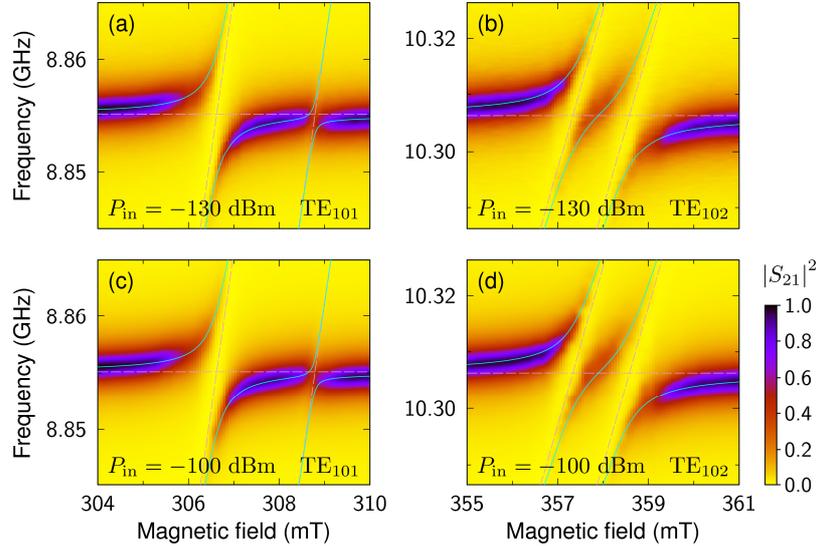

**Figure 2 | Strong magnon-photon coupling achieved at cryogenic temperature.** In **(a)-(d)**, the transmission spectrum of the rectangular 3D cavity with a small YIG sphere is measured as a function of the static magnetic field at 22 mK in a dilution refrigerator. The input microwave power is **(a)** −130 dBm with different frequencies around TE$_{101}$, **(b)** −130 dBm with different frequencies around TE$_{102}$, **(c)** −100 dBm with different frequencies around TE$_{101}$, and **(d)** −100 dBm with different frequencies around TE$_{102}$. The horizontal dashed lines denote the resonant frequencies of the cavity and the tilted dashed lines show the magnetic-field dependence of the frequencies of the FMR (left) and MS (right) modes. The green solid curves correspond to the energy levels of the magnon-polariton obtained by diagonalizing the Hamiltonian in equation (3).

To obtain damping rates of the FMR and MS modes, we further fit our experimental results with the calculated transmission coefficient of the microwave cavity containing a YIG sphere. In a standard input-output theory (see, e.g., ref. 23), this transmission coefficient can be written as

$$S_{21}(\omega) = \frac{2\sqrt{\kappa_i \kappa_o}}{i(\omega - \omega_c) - \kappa_{tot} + \sum(\omega)}, \qquad (4)$$

where $\kappa_i$ ($\kappa_o$) is the corresponding input (output) dissipation rate due to the coupling between the feed line and the cavity. The total cavity decay rate is given by $\kappa_{tot} = \kappa_i + \kappa_o + \kappa_{int}$, with $\kappa_{int}$ being the intrinsic loss rate of the cavity. The self-energy $\sum(\omega)$ contains contributions from both the FMR and MS modes: $\sum(\omega) = \frac{\tilde{g}_{FMR}^2}{i(\omega - \omega_{FMR}) - \gamma_{FMR}} + \frac{\tilde{g}_{MS}^2}{i(\omega - \omega_{MS}) - \gamma_{MS}}$, where $\omega_{FMR}$ is the angular frequency of the FMR mode, which has a damping rate $\gamma_{FMR}$, and $\omega_{MS}$ is the angular frequency of the MS mode with a damping rate $\gamma_{MS}$. Using $S_{21}$ in equation (4) to fit the experimental results (see Supplementary Material for the simulated transmission spectra), we can evaluate the damping rates $\gamma_{FMS}$ and $\gamma_{MS}$ as well as the total decay rate $\kappa_{tot}$ of the cavity. These evaluated parameters and the calculated cooperativity are summarized in Table 1. The experimental parameters show that $\tilde{g}_m > \kappa_{tot}, \gamma_m$, with a cooperativity $C_m = \tilde{g}_m^2/\kappa_{tot}\gamma_m > 1$ in most cases, indicating that the strong coupling regime is reached when both FMR and MS modes in the YIG sphere interact with the same cavity mode. In the experiment of Fig. 2, the measurements were implemented at 22 mK using a dilution refrigerator. This very low temperature corresponds to a negligible average thermal photon number in the 3D cavity (e.g., $\sim 1 \times 10^{-2}$ for the fundamental cavity mode). Accordingly, for the probe microwave tone in the measurements, we used a very weak input power of −130 dBm in Figs. 2a and 2b and a weak input power of −100 dBm in Figs. 2c and 2d, which correspond to average microwave photon numbers ~ 0.8 and 800, respectively, for the fundamental cavity mode. Note that the measured transmission spectra in Figs. 2c and 2d are very close to those in Figs. 2a and 2b, even though the average microwave photon number in the cavity has increased from ~ 0.8 to 800. Owing to the small number of photons in the cavity, especially in the case of −130 dBm, the measured results clearly show that a strong magnon-photon coupling has been experimentally achieved in the quantum regime for both the FMR and MS modes coupled to the same cavity mode. Moreover, because $\gamma_{MS} \sim 2\gamma_{FMR}$ at 22 mK (see Table 1), the experimental results reveal that the newly observed MS mode exhibits quantum coherence nearly as good as the FMR mode at cryogenic temperature.

Table 1: Summary of the experimentally obtained parameters

| Temperature | Mode | $\tilde{g}_{FMR}/2\pi, \tilde{g}_{MS}/2\pi$ (MHz) | $\kappa_{tot}/2\pi$ (MHz) | $\gamma_{FMR}/2\pi, \gamma_{MS}/2\pi$ (MHz) | $C_{FMR}, C_{MS}$ |
|---|---|---|---|---|---|
| 22 mK | $TE_{101}$ | 5.4, 1.4 | 1.1 | 1.2, 2.7 | 22.1, 0.7 |
|  | $TE_{102}$ | 7.5, 8.3 | 2.4 | 1.3, 3.3 | 18.0, 8.7 |
| Room | $TE_{101}$ | 5.2 | 2.5 | 1.3 | 8.3 |
|  | $TE_{102}$ | 9.6 | 5.9 | 1.5 | 10.4 |

**Experimental measurements at room temperature**

As an explicit comparison, we also measured, at room temperature, the transmission spectrum of the 3D microwave cavity containing the same YIG sphere (see Fig. 3). Now the average number of thermal photons becomes larger (e.g., $\sim 1 \times 10^3$ at 300 K for the fundamental cavity mode), so one has to use a probe microwave tone with higher power. In Fig. 3, the input microwave power is −20 dBm, which corresponds to an average microwave photon number $\sim 1.8 \times 10^{10}$ for the fundamental cavity mode. Similar to the experimental observations in ref. 14, at room temperature, an avoided crossing also occurs when the applied static magnetic field is tuned to make the FMR mode in resonance with the cavity mode $TE_{101}$ or $TE_{102}$. This reveals that the low-lying excitation condition $\langle b^\dagger_{FMR} b_{FMR} \rangle / 2S^{(FMR)} \ll 1$ is satisfied for the FMR mode even at room temperature, because the FMR-cavity Hamiltonian can be reduced to a form similar to equation (3), where the FMR mode is described by non-interacting magnons. This is due to the fact that the FMR mode involves the uniform precession of all spins, such that $S^{(FMR)} = Ns$, where the total number $N$ of spins in the YIG sphere is many orders of magnitude greater than unity. Moreover, based on the fitting results for the same YIG sphere, the damping rate of the FMR mode is found to slightly increase from 1.2 MHz (1.3 MHz), at ~ 22 mK, to 1.3 MHz (1.5 MHz) at room temperature, when the frequency of the FMR mode is close to the fundamental (second) cavity mode (see Table 1). These results reveal that the FMR mode is robust against temperature. The damping rate of the FMR mode obtained here is comparable to that in ref. 14, where cavity quantum electrodynamics with a YIG sphere of 0.36 mm in diameter was investigated only at room temperature. Note that the only magnetic ions in YIG are the ferric ions. Because these ions are in an $L = 0$ state with a spherical charge distribution, their interaction with lattice deformations and phonons is weak[16]. As a result, the FMR mode which involves the uniform precession of all spins is only weakly affected by phonons. This explains why the obtained damping rate of the FMR mode does not change much from cryogenic to room temperatures.

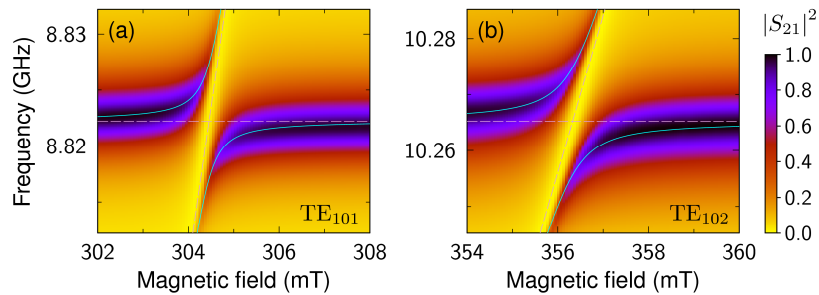

**Figure 3 | Strong coupling of the FMR mode to two cavity modes at room temperature.** The transmission spectrum of the rectangular 3D cavity containing the same YIG sphere is measured at room temperature as a function of the static magnetic field. The input microwave power is −20 dBm with different frequencies around **(a)** $TE_{101}$ and **(b)** $TE_{102}$, respectively.

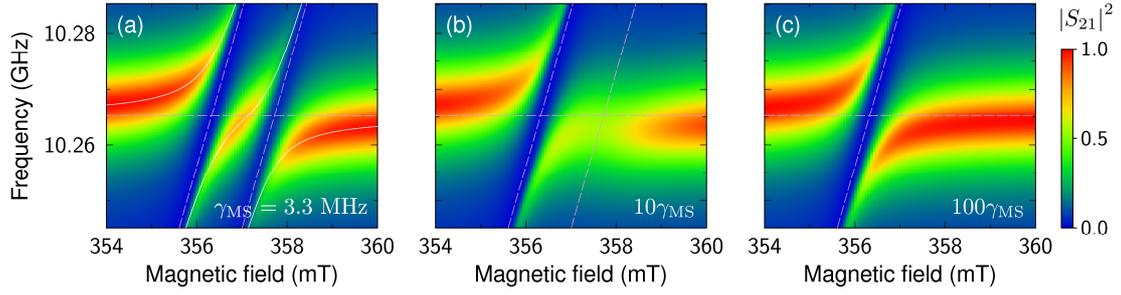

**Figure 4 | Impact of the MS-mode damping rate on the transmission spectrum.** In **(a)-(c)**, numerical simulations of the transmission spectrum are performed using equation (4) by choosing different input microwave frequencies around $TE_{102}$. The parameters for both the cavity and FMR modes are the same as the measured values in Fig. 3b. For the MS mode, the parameters are the same as the measured value in Fig. 2b, but its damping rate increases from **(a)** $\gamma_{MS}$ =3.3 MHz, to **(b)** $10\gamma_{MS}$, and then to **(c)** $100\gamma_{MS}$.

For the MS mode, the avoided crossing observed at cryogenic temperatures disappears at room temperature (comparing Fig. 3 with Fig. 2). This disappearance of the MS mode may be due to (i) the low-lying excitation condition for the MS mode is not satisfied at room temperature, and (ii) the strong magnon-photon coupling regime was not reached. When the frequency of the MS mode is close to the cavity mode, $\langle b_{MS}^\dagger b_{MS}\rangle \sim 1\times 10^3$ at room temperature. The low-lying excitation condition $\langle b_{MS}^\dagger b_{MS}\rangle / 2S^{(MS)} \ll 1$ is not followed if $\langle b_{MS}^\dagger b_{MS}\rangle \sim 2S^{(MS)}$. However, at cryogenic temperatures, the observed transmission spectrum shows that $\tilde{g}_{MS} \equiv g_{MS}\sqrt{2S^{(MS)}}$ is comparable to $\tilde{g}_{FMR} \equiv g_{FMR}\sqrt{2S^{(FMR)}}$ (see Table 1). Thus, $S^{(MS)} \sim S^{(FMR)}$, implying that for the MS mode, the low-lying excitation condition $\langle b_{MS}^\dagger b_{MS}\rangle / 2S^{(MS)} \ll 1$ is still satisfied even at room temperature. Therefore, the first possibility is ruled out for the disappearance of the MS mode at room temperature. In Fig. 4, we present numerical simulations of the transmission spectrum around the frequency of the cavity mode $TE_{102}$, by increasing the damping rate of the MS mode. It can be seen that when the damping rate $\gamma_{MS}$ is increased by two orders of magnitude, the simulated transmission spectrum agrees well with the experimental result obtained at room temperature (comparing Fig. 4c with Fig. 3b). This damping rate is much larger than the coupling strength $\tilde{g}_{MS}$, corresponding to a weak-coupling regime between the MS mode and the cavity mode. Therefore, in the transmission spectrum, the disappearance of the MS mode at room temperature can be attributed to the drastic increase of the MS-mode damping rate with temperature.

## Discussion

Note that the experimental results in ref. 13 show that the linewidth of the FMR mode increased more than three times when the temperature was raised from $\sim 10$ mK to $\sim 10$ K. This observed linewidth increase of the FMR mode with the temperature was ascribed to the slow-relaxation process due to the impurity ions and magnon-phonon scattering[13]. In contrast, the damping rate (i.e., the linewidth) of the FMR mode in our YIG sphere does not show an appreciable contribution from this slow-relaxation process, because $\gamma_{FMR}$ slightly increases when raising the temperature from 22 mK to room temperature (see Table 1). Thus, our YIG sample is of a better quality than that in ref. 13, regarding the quantum coherence of the FMR mode. In ref. 14, cavity quantum electrodynamics with a YIG sphere of 0.36 mm in diameter was studied at room temperature. The obtained damping rate of the FMR mode is $\sim 1.1$ MHz, which is also comparable to our result at room temperature. Moreover, for YIG doped with gallium, it was reported[24] that the linewidth of the FMR mode decreases from 1 mT at 4.2 K to 0.1 mT at room temperature, corresponding to 28 MHz and 2.8 MHz, respectively. This counter-intuitive observation of the linewidth narrowing for increasing temperature is different from both our results and those in ref. 13. Therefore, the linewidth of the FMR mode in the YIG material and the corresponding quantum-coherence behavior raise interesting open questions for further investigations. In our experiment, only measurements at cryogenic and room temperatures were implemented due to limitations in the experimental facility, so it is desirable (though challenging) to perform measurements at the wide-range intermediate temperatures, so as to provide more insights into the damping mechanism of the magnons.

In ref. 15, the ultrastrong coupling between the FMR mode and the bright mode of a reentrant cavity was observed at 20 mK by using a larger YIG sphere (~0.8 mm in diameter) attached to the reentrant cavity. In addition, a strong coupling between a MS mode and the dark mode of the reentrant cavity was also observed. In sharp contrast to our observations, the cavity mode coupled to the FMR mode and the cavity mode coupled to the MS mode were not the same mode of the reentrant cavity. Moreover, in ref. 15, the cavity-MS mode coupling was much weaker than the cavity-FMR mode coupling. This is also very different from our results and might be due to the different symmetries of the relevant magnon modes or the material properties of the YIG sphere.

In conclusion, we have experimentally studied cavity quantum electrodynamics with ferromagnetic magnons in a small YIG sphere at both cryogenic and room temperatures. We observed strong coupling of the same cavity mode to both FMR and MS modes in the quantum limit at ~22 mK. Also, we observed strong coupling of the cavity mode to the FMR mode at room temperature, with the damping rate of the FMR mode only slightly increased. This reveals the robustness of the FMR mode against temperature. Moreover, we find that the MS mode disappears at room temperature in the measured transmission spectrum and also numerically show that this is due to a drastic increase of the damping rate of the MS mode. Our work unveils quantum-coherence properties of the ferromagnetic magnons in a YIG sphere at both cryogenic and room temperatures, especially the extraordinary robustness of the FMR mode against temperature.

## Material & Methods

**Sample preparation.** The rectangular 3D cavity was produced using oxygen-free high-conductivity copper and its inner faces were highly polished. Also, this cavity was designed to have two connector ports for measuring microwave transmission. A commercially sold YIG sphere with a diameter of 0.32 mm was made by China Electronics Technology Group Corporation 9th Research Institute. This YIG sphere was mounted in the rectangular 3D cavity and the measurement of the microwave transmission through the cavity was implemented at both cryogenic and room temperatures.

**Measurement.** An Oxford Triton 400-10 cryofree dilution refrigerator (DR) was employed for the cryogenic measurement and a temperature as low as 22 mK was achieved at the mixing chamber stage of the DR. The cavity was placed in the mixing chamber stage and applied a static magnetic field using a superconducting magnet. In order to prevent the room-temperature thermal noise from the input lines, we used a series of attenuators to reach a total attenuation of 100 dB. For the output line, a circulator was attached to achieve an isolation ratio of 30 dB. Then, the transmitted signal from the cavity was amplified via a series of amplifiers both at the 4 K stage and outside of the DR. The microwave transmission measurement was performed using a vector network analyzer (VNA; Agilent N5030A, Agilent Technologies, Santa Clara, California, USA). To adjust the average photon number in the cavity, the input microwave power was tuned to be either −30 dBm or 0 dBm. Because of the total attenuation of 100 dB on the input lines, the final input power in the cavity becomes −130 dBm or −100 dBm.

For the room-temperature measurement, the same cavity with the YIG sphere was placed in a static magnetic field generated by an electromagnet (made by Beijing Cuihaijiacheng Magnetic Technology Co., Ltd., Beijing, China). The transmission spectrum of the hybrid system was measured using a VNA (Agilent N5232A) directly connected to two ports of the cavity with an input power of −20 dBm.

**Acknowledgement** This work is supported by the NSAF Grant No. U1330201, the NSFC Grant No. 91421102, and the MOST 973 Program Grant Nos. 2014CB848700 and 2014CB921401. F.N. is partially supported by the RIKEN iTHES Project, the MURI Center for Dynamic Magneto-Optics, the impact program of JST, and a Grant-in-Aid for Scientific Research (A).

# Supplementary Material

## 1. Measured transmission spectrum of the cavity without a YIG sphere

Our rectangular three-dimensional (3D) cavity with input/output ports was made of oxygen-free copper and has dimensions $50 \times 18 \times 3$ mm$^3$. By adjusting the lengths of the pins inside the input and output ports, the coupling rates of these two ports to the relevant cavity modes are tuned to $\kappa_{i,1}/2\pi = 0.19$ MHz and $\kappa_{o,1}/2\pi = 0.20$ MHz for the fundamental (first) mode TE$_{101}$ of the cavity, and $\kappa_{i,2}/2\pi = 0.85$ MHz and $\kappa_{o,2}/2\pi = 0.99$ MHz for the second mode TE$_{102}$ of the cavity. Figure S1(a) shows the transmission spectrum of the 3D cavity without a YIG sphere, as measured at 22 mK. It is found that the frequencies of the cavity modes TE$_{101}$ and TE$_{102}$ are $\omega_{c,1}/2\pi = 8.855$ GHz and $\omega_{c,2}/2\pi = 10.306$ GHz, respectively. The total cavity decay rates are measured to be $\kappa_{tot,1}/2\pi = 1.1$ MHz for the cavity mode TE$_{101}$ and $\kappa_{tot,2}/2\pi = 2.4$ MHz for the cavity mode TE$_{102}$, where the corresponding intrinsic loss rates of the cavity are 0.71 MHz and 0.56 MHz, respectively.

When the transmission spectrum of the 3D cavity without a YIG sphere is measured at room temperature, the resonance frequencies of the cavity shift and the corresponding intrinsic loss rates increase. This is due to the changes of the mechanical and material properties of the 3D cavity, as compared with the cavity at 22 mK. Figure S1(b) shows the measured transmission spectrum of the 3D cavity at room temperature. From the measurement results, it is found that the frequencies of the cavity modes TE$_{101}$ and TE$_{102}$ are $\omega_{c,1}/2\pi = 8.822$ GHz and $\omega_{c,2}/2\pi = 10.265$ GHz, respectively. The total cavity decay rates are measured to be $\kappa_{tot,1}/2\pi = 2.5$ MHz for the cavity mode TE$_{101}$ and $\kappa_{tot,2}/2\pi = 5.9$ MHz for the cavity mode TE$_{102}$, where the corresponding intrinsic loss rates of the cavity are 2.11 MHz and 4.06 MHz, respectively.

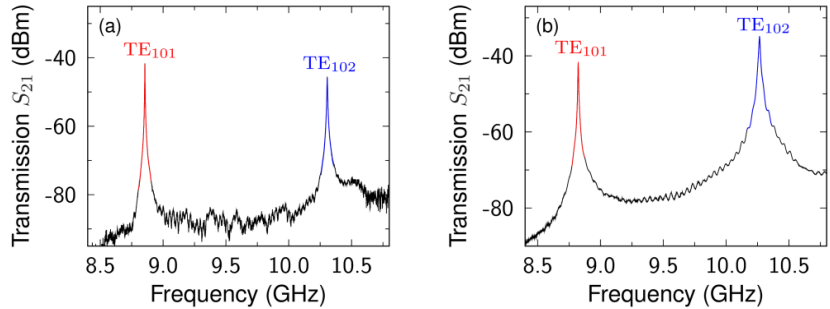

FIG. S1: **Transmission spectrum of the rectangular 3D cavity without a YIG sphere.** (a) Measured transmission spectrum at 22 mK. (b) Measured transmission spectrum at room temperature.

## 2. Calculated transmission spectrum of the cavity containing a YIG sphere

The transmission of the 3D cavity containing a YIG sphere can be calculated using equation (4) in the main text. Figures S2 and S3 show the calculated transmission spectra, so as to compare with the transmission spectra measured at 22 mK and room temperature, respectively. The parameters used in the simulations are all extracted from the measurement results as provided in the main text and Table 1. Comparing Fig. S2 (S3) with Fig. 2 (3) in the main text, one can see that the calculated transmission spectra agree very well with the measured ones. This indicates the validity of the extracted parameters from the measurement results.

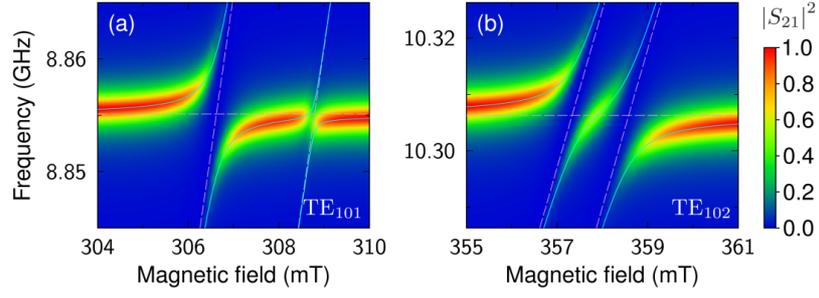

FIG. S2: **Calculated transmission spectra at cryogenic temperature**. Numerical calculations of the transmission spectrum are performed using equation (4) with extracted parameters in the main text and Table 1. The calculated spectra are displayed with different frequencies around (a) the first cavity mode $TE_{101}$ and (b) the second cavity mode $TE_{102}$.

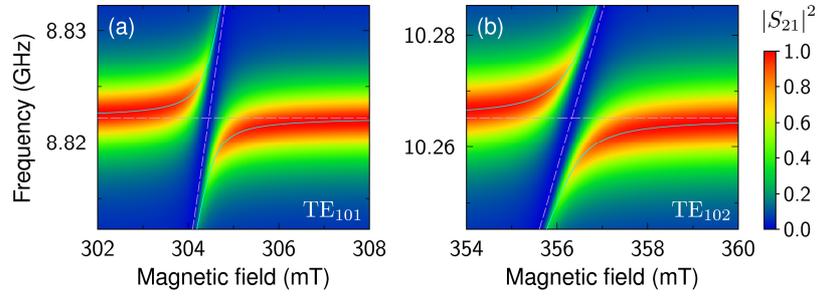

FIG. S3: **Calculated transmission spectra at room temperature.** The calculated spectra are also displayed with different frequencies around (a) the first cavity mode $TE_{101}$ and (b) the second cavity mode $TE_{102}$.